\documentclass[]{basi}
\usepackage{txfonts}
\usepackage{graphicx}
\usepackage{natbib}

\def\msun{M_\odot}
\begin{document}
\title[Problems of Collisional Stellar Dynamics]
{Problems of Collisional Stellar Dynamics}
\author[D.~C.~Heggie]%
       {D.~C.~Heggie$^1$\thanks{e-mail:d.c.heggie@ed.ac.uk}\\
%        A.~B.~Second$^{1,2}$ and C.~D.~Third$^2$\\
       $^1$University of Edinburgh, 
School of Mathematics and the Maxwell Institute for Mathematical Sciences, \\
King's Buildings, Edinburgh EH9 3JZ, U.K.}%% \\
%%        $^1$The Second Institude, Address, City, Country}

\pubyear{2010}
\volume{00}
\pagerange{\pageref{firstpage}--\pageref{lastpage}}
\date{Received \today}

\maketitle
%------------------------------------------------------------------------------%
% abstract and keywords                                                        %
%    see ?? for valid keywords                                                 %
%------------------------------------------------------------------------------%
\label{firstpage}

\setcounter{footnote}{1}
\begin{abstract}
The discovery of dynamical friction was Chandrasekhar's best known
contribution to the theory of stellar dynamics, but his work ranged
from the few-body problem to the limit of large $N$ (in effect,
galaxies).   Much of this work was summarised in the text ``Principles
of Stellar Dynamics'' \citep{C1942,C1960}, which ranges from a precise
calculation of the time of relaxation, through a long analysis of
galaxy models, to the behaviour of star clusters in tidal fields.  The
later edition also includes the work on dynamical friction and related
issues. %%  {\bf Does it also include the work on binaries?}
In this review we
focus on progress in the collisional aspects of these problems,
i.e. those where few-body interactions play a dominant role, and so
we omit further discussion of galaxy dynamics.\footnote{There is one
  other such problem to which Chandrasekhar contributed, though the
  paper in question \citep{C1944} was not reprinted in the book.  See
  Sec.\ref{sec:binaries}.  For more on the collisionless problems
  studied by Chandrasekhar, see the paper by N. Wyn Evans in the
  present volume.}  But we try to
link Chandrasekhar's fundamental discoveries in collisional problems
with the progress that has been made in the 50 years since the
publication of the enlarged edition.  
 
%% ........
%% This guide explains the usage of the \LaTeX\ class file for the Bulletin of the
%% Astronomical Society of India.  The class file \verb+basi.cls+ as well as this
%% guide are available electronically from the BASI webpage:\\
%% %
%% \verb+http://www.ncra.tifr.res.in/~basi/+.
%
\end{abstract}


%------------------------------------------------------------------------------%
% main text of the paper, using \section, \subsection, \subsubsection          %
%------------------------------------------------------------------------------%

\section{Introduction}\label{s:intro}

Chandrasekhar's ``Principles of Stellar Dynamics'' is not his
best-known text, but it had few competitors for many years, and
covered a broad range of topics.  The later edition \citep{C1960}
added a number of lengthy and significant papers mainly on the
statistical approach to collisional stellar dynamics, and was
published just over 50 years ago.  In this review we consider a few of the topics considered by Chandrasekhar, and try to connect
his view of the subject with current research.  

Since the book is not so well known, and has been virtually supplanted
by the book by Binney and Tremaine \citep{BT1987,BT2008}, it is worth
looking over its contents list.  After an observational review, the
theory
begins with a  careful derivation of the relaxation time of stellar systems,
including all the geometry of two-body encounters and a discussion of
the origin of the Coulomb logarithm.  It contains a derivation of a
formula for what came to be known as ``dynamical friction'', possibly
Chandrasekhar's most significant and enduring discovery in the field
of stellar dynamics.  The history and context of Chandrasekhar's work
in these two topics is nicely discussed in \citet{P1996}, and in the
present paper we
consider more recent developments in the theory of dynamical friction
in Sec.\ref{sec:df}.

In Chandrasekhar's book there follow two chapters on collisionless
stellar dynamics, or rather the dynamics of galaxies.  After some
preliminaries, the first of these considers the problem of determining
what galactic potentials are consistent with the assumption of a
Schwarzschild (Gaussian) distribution of velocities, while the second
turns to the problem of spiral structure.  Then there is a long and
interesting chapter on collisional stellar dynamics; specifically, the
dynamics of star clusters, a subject which we review here in Sec.\ref{sec:dsc}.

Apart from two appendices, at this point the old and new editions of
the book differ.  The latter now includes several reprints, some on
dynamical friction (a topic we take up here in Sec.\ref{sec:df}), and
a final long paper on the statistics of the gravitational field of a
distribution of point masses, together with its applications to
dynamical friction and star clusters.  It is amusing to find that
Chandrasekhar titled this last paper ``New Methods in Stellar
Dynamics''.  One wonders if this was a conscious echo of Poincar\'e's
famous ``Les M\'ethodes Nouvelles de la M\'ecanique Celeste'' (see also \citealt{H1967}).  At any
rate, one of Chandrasekhar's applications of his theory is the starting
point for the next section of this review.

%**********{\bf complete this}

%................edited to here 31/12/10

%------------------------------------------------------------------------------%
\section{The Dynamics of Binary Stars}\label{sec:binaries}

We begin with a  slim paper ``On the Stability of Binary Systems''
\citep{C1944}.  It did not make it into his book, but it 
appears to be Chandrasekhar's only work on a  topic which has become
one of the pillars of collisional stellar dynamics.  In
Chandrasekhar's paper it is set in the context of a critique  by
Ambartsumian \citep{A1937} of an older idea by Jeans, who had used information on
the distribution of binary stars to argue that the Milky Way was well relaxed.  

\subsection{The statistical effect of encounters}\label{sec:encounters}

Chandrasekhar's estimate for the disruption time scale, $\tau$, of a
binary  was based on his theory of the two-point
distribution for the gravitational acceleration due to a random
distribution of stars, which led to the formula
$$
\tau = \frac{(M_1+M_2)^{1/2}}{4\pi G^{1/2}MNa^{3/2}},
$$
where $M_1,M_2$ are the component masses, $a$ is the semi-major axis,
and $M,N$ are the average individual mass and number density of the
field stars.  Ambartsumian's formula, by contrast, was 
$$
\tau = \frac{v}{4\pi GMaN\ln\left(1+ \displaystyle{\frac{a^2v^4}{4G^2M^2}}\right)},
$$
where $v$ is some average speed which we take here to be the velocity dispersion.
Notice that, by contrast, Chandrasekhar's formula includes no
information on the velocity dispersion, because the underlying theory
describes the spatial correlation of fluctuations but not their
temporal correlation.  He gives the impression that the absence of any
dependence on the velocities is a merit, but it later turned out that
the velocities of the stars are
crucial.  Jeans himself \citep{Jeans1918} had argued (incorrectly, as
it later emerged) that encounters with field stars would lead to
equipartition of kinetic energies, giving all binaries a period of
order
\begin{equation}
P = \frac{G(M_1+M_2)}{v^3}.\label{eq:gl}
\end{equation}  

This conclusion was, however, turned upside down by \citet{GL1950}, who used arguments akin to those of Ambartsumian and
obtained formulae for the average rate of change of the binding energy
of a binary as a result of encounters.  They concluded that, if a
binary has period much longer  than eq.\ref{eq:gl}, then its period
will tend to become longer still (eventually leading to disruption), while if
its period is much shorter then it becomes still shorter.  Their
conclusion was correct, and was arrived at independently by Hills
\citep{AH1972,Hills1975} using numerical methods and by Heggie
\citep{H1975} using mainly analytical approximations.  Heggie also
used the terms ``hard'' and ``soft''  to signify binaries whose internal binding energy, $\varepsilon$, was larger or smaller,
respectively, than the mean kinetic energy of the field stars.

The case of equal masses has been worked out in a lot of detail,
especially in a series of papers by Hut and colleagues, much of it
summarised with tables, figures and formulae in \citet{HH1993}.
Applications, of course, require unequal masses in general, and here
our knowledge is much more patchy.  \citet{HHM1996} were able to give
a general formula for exchange encounters with hard binaries
(i.e. encounters in which the incoming third star takes the place of
one of the original binary components).  They used 
analytical arguments to establish the mass dependence for extreme
cases (e.g. one component of very low mass), and filled in the gaps by
interpolation in results of numerical  experiments.  For
this purpose they used the starlab package
(http://www.manybody.org/manybody/starlab.html, \citet{MH1996}), 
which has very well
organised tools for the computation of scattering cross sections.
Large numbers of other results for various specific combinations of masses
will be found scattered in the literature, including \citet{SP1993}
and especially the compendious book of \citet{VK2006}.

A number of extreme parameter ranges have become important for
astrophysical reasons, and also these are situations in which the
complexities of the mass-dependence of a cross section may be
considerably reduced.  On the other hand, as we shall see, the
situation can be considerably richer than the simple notion that soft
binaries soften and hard binaries harden.

The
case of a third body (intruder) of relatively low mass has been
studied in the context of the hardening of a black hole-black hole
binary in a galactic nucleus \citep{MV1992}.  It led to an interesting
debate \citep{Hills1990,Gould1991} on whether it is really true that
hard binaries (defined by the ratio of the binding energy of the
binary and the energy of relative motion of the intruder) tend to
harden and soft binaries tend to soften.  Hills had argued that it was
the ratio of {\sl speeds} that mattered, but \citet{Q1996} eventually
vindicated the earlier position.  There is now a considerable
literature on the problem of a massive binary in a system of particles
of low mass which use $N$-body simulations rather than cross sections.

Another important context where the distinction between fast
encounters and energetic encounters is significant is the study of
stellar encounters with planetary systems.  This has been studied by
several groups, including \citet{LA1998,HD1989,MDH2010,SGHL2009}, but
here we focus briefly on the study of \citet{FCR2006}.  In the case
under consideration, let us denote the incomer velocity which distinguishes hard
from soft binaries (in the sense of energies) by $v_c$ (the
``critical'' velocity, in the sense that slower encounters cannot
destroy the binary).  Because the planetary mass is so low, $v_c$ is
much smaller than the orbital speed of the planet ($v_{orb}$).  These
authors find that, indeed, when the encounter speed is less than a
speed of order $v_c$, the average change in the binding energy of the
 binary is positive, i.e. the encounter hardens the binary.
But at the same time the planetary system has been destroyed, as the
most likely outcome of a close encounter in this regime is an exchange
encounter leaving the two stars bound.  Similarly, in the regime 
of encounter speeds between $v_c$ and $v_{orb}$ an encounter indeed tends to soften
a planetary system, but not to disrupt it (until encounter speeds of
order $v_{orb}$ are reached).  A careful reading of \citet{FCR2006} is
recommended for a proper appreciation of the issues.

The last case of extreme masses that we shall mention is another
highly topical one: that of a single black hole encountering a binary
with stellar-mass components.  This is thought to be of importance in
the creation of high-velocity stars by scattering off the black hole
at the Galactic Centre \citep{Hills1988}.  The literature is
considerable, but among those studies focusing on the three-body
aspects of the problem are
\citet{ZLU2010,GGPZ2009,SKR2010,GPZS2005,YT2003}.

\subsection{Wide binaries in the Milky Way}

Chandrasekhar's interest in binary stars was focused on the
dynamical evolution of field binaries, a topic which remains lively
up to the present day, with an extensive literature, especially on the
observational side.  Before turning to recent developments, however,
it is worth mentioning that the topic had already been considered,
before the work of Chandrasekhar and Ambartsumian, by \citet{Op1932}.  

This thread of research was then taken up by \citet{O1950} who, like
\"Opik, was concerned with binaries with one massless component (a
comet or meteoroid).  While all these studies used analytical
estimates, soon after this numerical integrations came into routine use,
and were applied to this problem by \citet{Ya1966} and \citet{CP1971}.
The latter authors found that the lifetime exceeded the estimates
given by any of the previous theories which they tested.  There were
also theoretical developments, however.  Except for Chandrasekhar's
theory, that earlier work had been based on the computation of the
mean square change of velocity (i.e. the relative velocity between the
two components of a binary), or the mean change in the energy of the
binary.  As Chandrasekhar himself would have recognised, however, it
is also necessary to take into account the second moment of the change
in energy (i.e. its mean square value), to construct a kinetic theory
based on a Fokker-Planck treatment.  This was accomplished in
\citet{K1977} and \citet{RK1982}.

Further theoretical developments have mainly involved improvements in
the physical model, i.e. the inclusion of significant additional
processes, such as encounters with giant molecular clouds
\citep{WSW1987} and dark matter particles \citep{WW1987}.  Nor are the
wide binaries of the Galactic field lacking interest even after they
have dissolved.  Then they are also strongly subject to galactic
perturbations, which imposes an interesting (and potentially
detectable) correlation in density with a peak when the two components
have separated by 100-300pc \citep{JT2010}.  Finally, it is not
self-evident how wide binaries can emerge from the relatively dense
star-cluster environment in which most stars are thought to form, and indeed it
seems likely that significant numbers form during the cluster
dissolution process itself \citep{Ko2010}.

\section{The dynamics of star clusters}\label{sec:dsc}

The title of this section is also the title Chandrasekhar chose for
the last chapter of his book.  As usual, it opens with a number of
generalities, but then it moves on to the important problem of the
escape rate from a star cluster, including the differential escape of
stars of low and high mass.  Implicit in this theory is the assumption
that the cluster is isolated, but the next section of his book moves
on to consider the effect on a cluster of its galactic environment.
This section begins with an excellent derivation of equations of
motion, ``energy''-integral and virial theorem.  

\subsection{Tidal stability}\label{sec:tidal}

After the preliminaries, Chandrasekhar takes up the stability of star
clusters, using as his model an ellipsoidal cluster of uniform density
$\rho$.  The reason for this is that the gravitational field
(including the tidal field of the galaxy) becomes linear, and the
motions of the stars can be computed explicitly.  The frequencies
become imaginary when $\rho$ is sufficiently low, and Chandrasekhar
interprets this as the onset of instability.  A somewhat similar
approach was taken by \citet{ACG1983}, who studied orbits in the
nearly constant-density core of a cluster.  They used Floquet analysis
to determine the stability limit and were thus able to extend results
to the case of a cluster on a non-circular galactic orbit.

This section of Chandrasekhar's book is of particular interest to the
author of the present paper because it turns out to be possible to
construct a self-consistent ellipsoidal model of uniform density by
superposition of these exact orbits \citep{MH2007}, though they are
indeed unstable when the density is low enough \citep{FH2005}.  Though
these models are artificial\footnote{The later paper has never been
cited so far, much as the referee predicted.  The present paper will
probably provide its one and only citation.}, they are of interest
because examples of self-consistent cluster models in a tidal field
are rare \citep{HR1995,BV2008,VB2009}.  These models also give a
pointer for the construction of better models of star clusters than
any in existence, in the following sense.  These models consist of the
familiar galactic epicycles, but modified by the attraction of the
cluster.  They are therefore retrograde orbits, and it has been known
for a long time that there should exist {\sl stable} retrograde orbits
in the vicinity of a star cluster, but outside its tidal radius
\citep{H1970}.  Thus one can imagine a sequence of self-consistent
cluster models with varying proportions of stars inside and outside
the cluster tidal radius, with (say) the models of \citet{HR1995}
(which are generalised King models) at one end of the sequence, and
the models of \citet{MH2007} at the other.

\subsection{Fokker-Planck dynamics}

Towards the end of his chapter on star clusters, in which
Chandrasekhar has discussed both escape and relaxation, he laments,
``A rigorous theory of galactic clusters must therefore take both
these factors into account.  But such a theory is not yet available.''
It was not too long in coming, the essential formalism being established
by \citet{Ku1957}.  But its power was first demonstrated by
\citet{H1961}, in a landmark paper which, among other things, produced
a solution of the Fokker-Planck equation (for the relaxation) with a
tidal boundary condition (producing escape).  Not only this, but
H\'enon also realised the critical role played by binaries.

H\'enon's model was of a very special type, but one which all
reasonable solutions would approach asymptotically.  It took almost
another 20 years before a general numerical solution of the equation
became feasible \citep{C1979}, though initially restricted to the case
of stars of equal mass, as in H\'enon's model.  This refinement was
added relatively quickly, however \citep{M1981,M1983}, albeit in the
context of {\sl galaxy} clusters (as opposed to galactic ones).  The
subsequent development of this tool was steady: rotation
\citep{G1983}, binary stars formed in three-body encounters
(quoted in \citealt{C1985}) or those formed tidally \citep{SOC1987} or primordially
\citep{G1991}and stellar evolution \citep{KCM1991} were all added;
until it became a tool which could be applied to the modelling of
individual star clusters and quite detailed comparison with
observations.

\begin{figure}
\includegraphics{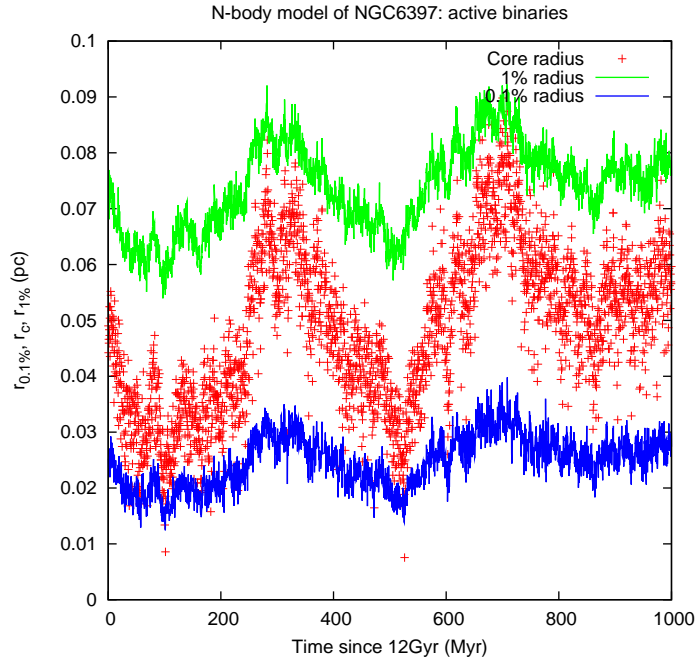}
\caption{Evolution of the core in a direct $N$-body model of the globular
  cluster NGC6397 (from \citet{HG2009}).  From top to bottom, the
  plotted radii are the 1\% Lagrangian radius (i.e. the radius of a
  sphere, centred at the densest part of the cluster, which contains
  1\% of the total mass), the core radius (i.e. the radius at which
  the density of the cluster falls to a certain fraction of its
  central value, though it is actually calculated by a different
  procedure), and the 0.1\% Lagrangian radius.  Radii are given in
  parsecs, and the horizontal axis is time (in units of 1Myr) after
  the present.  The cluster core is alternately compact and more
  extended, on a time scale which is long by comparison with the
  central relaxation time; this, and other arguments, suggest that the
  oscillations are essentially gravothermal.}\label{fig:gto}
\end{figure}

This was not the first kind of code to reach this goal, however.  In
advance of  the development of Fokker-Planck methods was a method of
treating the dynamics of a star cluster as if it was a
self-gravitating gas \citep{L1970}.  This technique developed with
comparable rapidity, and after only 10 years it was possible to produce
synthetic surface density profiles for comparison with observation
\citep{ADG1980}.  These gas models remained of importance, and were
responsible for the discovery of gravothermal oscillations
\citep{SB1983}, which are the response of a system to an unstable
balance between the relaxation-driven flow of energy and the formation
of energy by binary interactions in the core.  Interest in this
behaviour slowly waned, but has recently been invoked as a significant
mechanism for understanding the variety of surface brightness profiles
exhibited by well observed star clusters (Fig.\ref{fig:gto}).

\subsection{Monte Carlo models}

From the numerical point of view, both the Fokker-Planck and gas
models are of finite difference type.  It is also possible to solve
the former equation with at least two kinds of Monte Carlo technique.
One of these was pioneered by Spitzer and his students (see
\citet{SH1971} and subsequent papers in the series).  Its last serious
application appeared many years ago \citep{SM1980}, and it is probably
ripe for revival, as it better adapted to some important situations
(e.g. a time-dependent tide) than some competitors.

An alternative Monte Carlo technique was developed at about the same
time \citep{H1967,H1971}, but has been taken up and developed by
others, right up to the present \citep{S1982,G2006,Ch2010}.  It now
includes a rich mix of important ingredients, not only relaxation and
escape, but also the internal (stellar) evolution of single stars and
binaries, as well as interactions involving binaries.  Despite its
limitations to a steady tidal field, spherical symmetry and zero
rotation, it is the method of choice for studying virtually all
globular star clusters, because it is so fast, without sacrificing
much realism.  Even the evolution of
a rich star cluster like 47 Tuc, which is thought to have almost
$2\times10^6$ stars and a few percent of binaries, can be modelled for
a Hubble time in less than a week on an ordinary computer
\citep{GH2010}.  Such modelling makes possible a range of
investigations at the interface with observations, and is very useful
for the planning and interpretation of some kinds of observational
programmes, such as searches for radial velocity binaries
\citep{S2009}.  The speed is important, because we do not know {\sl ab
  initio} what initial conditions to use to match a given cluster.
Repeated trial and error, or grid searches, require a fast method.

\subsection{$N$-body methods}\label{sec:nbody}

Naturally enough, there is nothing in {\sl Principles of Stellar
  Dynamics} which prepares us for the explosion of interest in $N$-body methods in
  the subject, even if we restrict ourselves to direct summation
  methods.  It started about 20 years after the publication of the
  book \citep{vH1960}, and in 50 years has brought us to the point where it begins to
  be possible to study the entire life history of the easiest globular
  clusters \citep{Z2011}, though most still lie beyond our capabilities.

\begin{figure}
   \centerline{\includegraphics[angle=0,width=12cm]{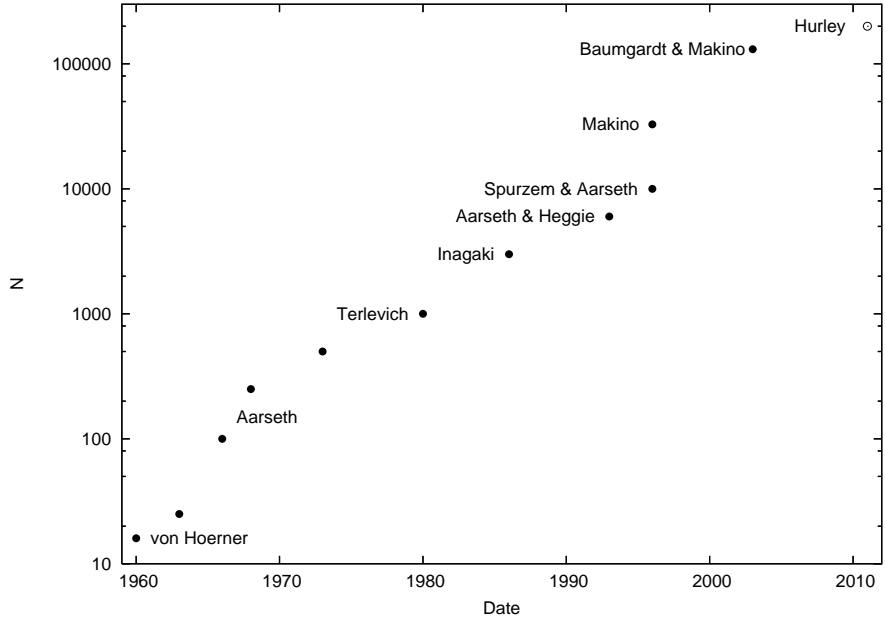}}
   \caption{Largest direct $N$-body simulation to date, plotted
   against publication date.  Only dynamically well evolved
   simulations (roughly speaking, to or beyond ``core collapse''; see, for example, \citet{HH2003}) are
   included.  The last simulation is not yet published at the time of writing (February 2011), in fact (see \citet{Hu2008}).}  \label{fig:nbody}%
   \end{figure}

The main problem is the number of stars, $N$.  Fig. \ref{fig:nbody}
shows the steady but slow progress that has been made since 1960.  The
mean mass of the Galactic globular clusters being \citep{MSS1991} of order
$1.9\times10^5\msun$ (and the median mass is lower still), it might be
thought that a large fraction of them are within reach of $N$-body
simulation.  However, they lose large amounts of mass through
evolution over about 12Gyr, and so, except for a few sparse and large
clusters, the initial stages of the evolution prevents them from being
 simulated in a
reasonable time.  

Actually, it is not hard to evolve larger models than those shown in
Fig. \ref{fig:nbody} to well beyond core collapse.  Fig.\ref{fig:m4nb}
shows a simulation using as initial conditions those suggested for the
cluster M4 by \citet{HG2008}, except that there are {\sl no primordial
  binaries}.  If these had been included (and the suggested abundance
is only about 7\%) the progress of the
simulation would have been slower by about a factor of 20.

   \begin{figure}
   \centerline{\includegraphics[angle=0,width=12cm]{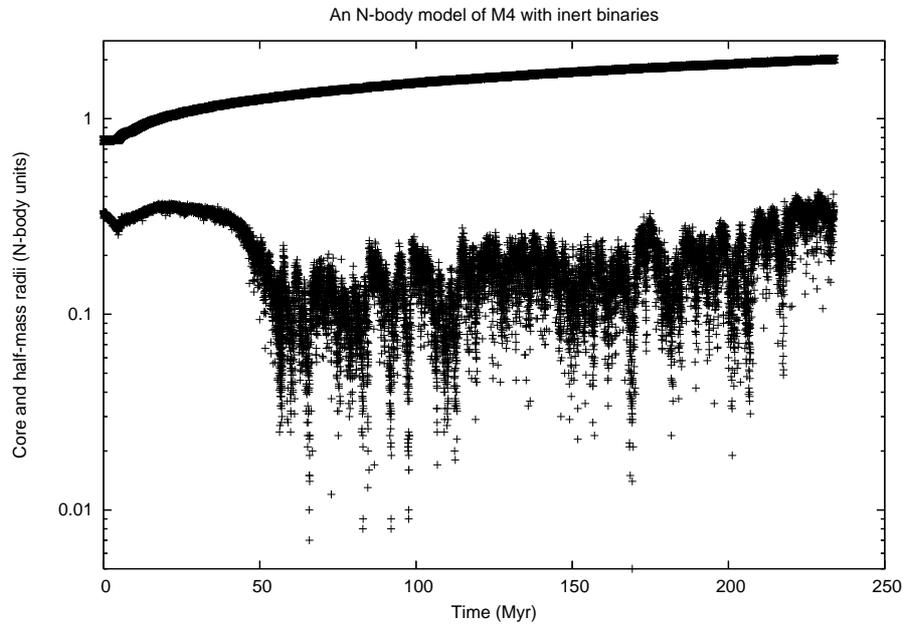}}
   \caption{Core- and half-mass radii of an $N$-body model with 453 000 particles initially.  The
     initial conditions are described in the text.  The model includes
     stellar evolution and the effect of the Galactic tide, but {\sl
     no binaries} (initially).  The initial brief reduction of the core radius is caused
     by mass segregation.  After about 5Myr, stellar evolution
     temporarily halts the collapse of the core, but this resumes, and
     reaches completion after about 50Myr.  Thereafter the small core
     is sustained by the formation and dynamical evolution of binary
     stars, but the core is now unstable to ``gravothermal
     oscillations''.  Both stellar evolution (or rather its associated
     mass loss) and the evolution of binaries increase the energy of
     the cluster, which leads to the expansion of the half-mass
     radius.  The tidal radius (not shown), which marks the effective
     boundary between motions dominated by the cluster and those
     dominated by the Galaxy, decreases by only about 15\% in the time
     shown.  The units of radius are ``$N$-body units'' (see \citet{HH2003}).
This simulation took about 2 months.
}   \end{figure}\label{fig:m4nb}

\subsection{The escape rate from star clusters}

\citet{Ch1943a,Ch1943b} produced two papers on this topic in quick
succession.  His motivation for this was not simply to understand the
lifetime of star clusters, but to elucidate the role played by
dynamical friction (Sec. \ref{sec:df}).  Dynamical friction is an
aspect of two-body relaxation which tends to reduce the energy of
stars, especially those of high speed, and which therefore
particularly affects escaping stars.  Without it, Chandrasekhar
showed, the lifetime of a star cluster would be too short to explain
the existence of star clusters with ages of order a Gyr.  Somewhat
analogously, it has also 
been invoked in studies of the escape of black holes from a galaxy \citep{K1985a,K1985b}.

In Chandrasekhar's  papers he  was, in effect, solving the
Fokker-Planck equation assuming that escape took place at some fixed
speed (which he estimated from the virial theorem).  Similar
calculations have been carried out by \citet{SH1958,K1960} and \citet{LC2010}.
Long ago, however, \citet{K1958} pointed out a number of shortcomings
of the Chandrasekhar model, and proceeded to investigate some of them
in subsequent papers.  One of these was the spatial inhomogeneity of
the star cluster model, which, in the same paper, he investigated by
integrating the escape rate over the cluster.  The escape rate formula
which he integrated was a more primitive estimate than
Chandrasekhar's, dating back to the earlier
work of \citet{A1938} and \citet{S1940}.  A similar treatment, but
based on more elaborate  formulae for the local escape rate  were later presented by
\citet{Sa1976} and \citet{J1993}.

All the formulae in the papers cited are based on the theory of
relaxation, and therefore include as a factor the Coulomb logarithm.
A completely different view of the situation was taken by
\citet{He1960}, who showed that relaxation cannot lead to escape from
an isolated cluster, essentially because the ``period'' of a star's
orbit tends to infinity as its energy approaches zero (from below).
He obtained a formula for the escape rate due to discrete, individual
encounters (rather than the diffusive effective of many encounters).
Tellingly, it does not contain the Coulomb logarithm.  

One of the comments made by \citet{K1958} was that, however one
computes the escape rate, it will change as the cluster evolves.
\citet{SS1972} pointed out that relaxation changes the distribution
function of the stars in a cluster, and then a single encounter, as
envisaged by H\'enon, may raise its energy above the energy of
escape.  Therefore it is possible that the relaxation time scale does,
after all, control the escape rate from an isolated system, as is
commonly assumed.  

There have been many numerical studies addressing the problem, but we
shall mention only one \citep{BHH2002}, which showed that another,
additional mechanism is at play.  While it is true that most escapers
emerge from encounters deep inside the cluster, some occur because
this process causes the potential well of the cluster to become
shallower, and this in turn causes stars with energies just below the
energy of escape to drift across the escape boundary.

All these issues change when one considers a cluster limited by the
tidal field of a galaxy.  At the simplest level, the energy of escape
drops and the rate of escape increases, as was found numerically many
years ago \citep{W1968,Ha1970}.  But the very notion of escape is
complicated.  It is possible, at least on the standard model of a star
cluster on a circular galactic orbit, for an escaper to recede
arbitrarily far from from the cluster and still return to it
\citep{RMH1997}.  Stars can exist on stable orbits with energies above
the escape energy, and even on orbits which lie outside the
conventional tidal radius of the cluster \citep{H1970}; see also
Sec.\ref{sec:tidal}.  Such stars have important effects on the
observable velocity dispersion profile of a globular cluster
\citep{KKBH2010}.  Matters are complicated further for clusters on
{\sl non-circular} galactic orbits, where there is no conserved
quantity analogous to energy (and therefore no notion of escape energy
or even of an escape boundary).  Nevertheless the common view is that,
even in these cases, the time scale of escape is determined by the
time of relaxation.

Among early indications that this is not so were numerical results by
\citet{VH1997}, who showed that the escape rate depended
systematically on $N$, even when scaled by the relaxation time.
$N$-body models by T. Fukushige and J. Makino \citep{HGST1998} showed
clearly that escape scales with $N$ in a different way from
relaxation.  Further $N$-body results \citep{He2001a,He2001b} gave an
escape time scale proportional (empirically) to about $N^{0.63}$,
whereas in the same units the relaxation time scales approximately as
$N/\log N$.

The problem was greatly clarified by the work of \citet{Ba2001}.  He
showed that the scaling could be understood by noting that stars may
remain inside a static cluster for an arbitrarily long time, even with
energies above the escape energy \citep{FH2000}, and that this changes
the escape time scale from the relaxation time, $t_r$, to
approximately $t_r^{3/4}t_{cr}^{1/4}$, where $t_{cr}$ is the crossing
time.  For the range of $N$ studied in $N$-body simulations of the
time, this results in a dependence close to $N^{0.63}$, in units such
that $t_{cr}$ is constant.

It is the
interaction between this buffer of ``potential escapers'' and the
processes of relaxation and escape which complicate the overall escape
time scale.  The effect of this buffer is sometimes referred to as a
``retardation effect'', after a study by \citet{K1959}, which was in
turn suggested by a remark of \citet[p.209]{C1960}.  The point is that
a star which has gained enough energy to escape may, on its way out of
the cluster, experience another encounter which brings it below the
escape energy once more.  But the $N$-dependence of this effect is
different from that described in \citet{Ba2001}, and there it is not
encounters which retain an escaper, but the dynamics of stars in the
field of tidal and inertial forces.

The scaling does depend 
on factors such as the initial concentration of the cluster
\citep{TF2005,TF2010}, the extent to which the cluster initially
underfills its tidal radius \citep{GB2008}, and the galactic environment
(i.e. the strength of the tidal field, \citet{LGP2005}).  Amazingly,
it does not depend significantly on the assumption that the cluster
orbit is circular \citep{BM2003}; if the orbit is non-circular, the
cluster appears to behave like one on a circular orbit of intermediate
radius.  Understanding this fact from a theoretical point of view 
is a significant unsolved problem in
this area, despite some empirical advances with the aid of
$N$-body simulations \citep{KKBH2010}.  There is also growing
observational evidence on the mass-dependence of cluster disruption,
and it is consistent with these theoretical developments
\citep{BL2003,GBLM2005,LG2006}, though it has to be recognised that
other processes come into play beyond the relatively gentle
evaporation of escapers created in encounters.  That is a long and old
story which we shall not review here.  

Equally old and long is the
theory of what is called {\sl preferential} or {\sl differential}
escape, i.e. its dependence on the mass of the escaper.  The common opinion
is that the escape rate increases with decreasing mass, but 
Chandrasekhar's finding \citep{C1960} (p.209f) was more subtle.  His
result was that the escape rate is fastest for stars with a mass of
about 40\% of the mean mass.  This result was based on the assumption
that stars are in energy equipartition in the cluster, which is
inconsistent with a fixed escape velocity.  Nevertheless the result
still turns out to hold in star clusters which include stellar
evolution and a low remnant retention fraction \citep{Kr2009}, if the
total disruption time of the cluster is short enough.

\section{Dynamical friction}\label{sec:df}

This is another subject with a long and rich history, and it takes us
beyond star cluster dynamics into the dynamics of galaxies and galaxy
clusters.  In that context, which we come to at the end of this
section, it also takes us away from the collisional problems to which
this review has been devoted.  Within collisional stellar dynamics,
dynamical friction 
is simply part of the mainstream of the theory of relaxation, and not often
receives separate, explicit mention.  There is an interesting
experimental check of what is, in effect, the coefficient of dynamical
friction in \citet{Th1996}.  Within limits the comparison is
satisfactory, but this study also shows that direct comparison is not
an easy task.

One current problem of collisional stellar dynamics involving
dynamical friction explicitly (and linking with the topics of section
\ref{sec:binaries}) is the fate of black holes in merging galaxies.
Their evolution was outlined in a famous paper of \citet{BBR1980}, and
much subsequent attention has been paid to a protracted period of
evolution under the action of dynamical friction, often referred to as
``the final parsec problem''.  Some theoretical studies relevant to
this problem (scattering of low-mass objects off a massive binary) are
referred to in Sec.\ref{sec:encounters}, and others which refer
specifically to the galactic context include \citet{PR1994} and
\citet{VCP1994}.
(Of course  it is a big assumption to suppose that this process
of the evolution of pairs of black holes can be
understood entirely in terms of stellar dynamics; the effect of
galactic gas and accretion disks around the black holes may be
decisive; but we shall continue to ignore these in our further review.)

In the stellar dynamical problem Chandrasekhar's formula has been
extended in several ways, e.g. to a non-uniform background medium
\citep{JP2005}, one with an anisotropic velocity distribution
\citep{Id2002}, or one with a mass spectrum \citep{Ci2010}.  $N$-body techniques are possible, but
demanding, because it is known on theoretical grounds that it is
necessary to include the effect of the black holes on the stellar
distribution self-consistently \citep{Iw2010,Se2010}.  To reach a
regime which can be scaled robustly to galactic nuclei is a
computational challenge comparable to the simulation of globular
clusters (Sec.\ref{sec:nbody}).  Progress has been faster than for the
globular cluster problem, however, partly because there is no need (it
is assumed) to follow also a binary population in the stellar
distribution \citep{BMS2005,Be2006,Be2009}.

Black holes are point masses, but the notion of dynamical friction has
been extended (in numerous studies) to problems of the orbital
evolution of a satellite galaxy within a larger galaxy or halo,
i.e. to extended bodies. From the theoretical point of view it seems
clear that the behaviour of a rigid satellite (which is the basis of
some theoretical studies) may differ essentially from that of a
responsive satellite \citep{FFM2006}.  A common approach is to use a
more-or-less self-consistent $N$-body simulation and to summarise the
results by a calibration of the Coulomb logarithm in the  Chandrasekhar
formula; examples include \cite{CMG1997,CMV1997,SFP2003} and \citet{Ju2010}.

While dynamical friction, as introduced by Chandrasekhar, is a
mechanism of collisional stellar systems, galaxies are collisionless
(at least, in the regime under discussion here).  Indeed, since
Chandrasekhar's time, it has become clear that there is a {\sl
collective} process which governs such phenomena as the decay of the
orbit of a satellite galaxy in the halo of its parent galaxy
\citep{TW1984,We1986,CP1998}.  It might even be concluded that the
physical phenomenon which causes the orbital decay of satellite
galaxies has no deeper connection with dynamical friction than the
same dependence on the basic scales of density and velocity dispersion
(which allows it to be expressed by a suitable choice of the Coulomb
logarithm).  Even more tenuous is the link between Chandrasekhar's
theory and the decay of a satellite in a partly or purely {\sl
gaseous} medium, though this too is often referred to as ``dynamical
friction''.  While there is some danger of confusing the underlying
physics, perhaps  it is a measure of the appeal of Chandrasekhar's
discovery that the term ``dynamical friction'' has been extended to
encompass such a diversity of astrophysical processes.

%------------------------------------------------------------------------------%
\section*{Acknowledgements}

I am very grateful to Don Goldsmith for permission to place his
translation of \citet{A1937} in the public domain. 

%------------------------------------------------------------------------------%
% bibliography:                                                                %
%------------------------------------------------------------------------------%

%\begin{thebibliography}{99}

\label{lastpage}
%------------------------------------------------------------------------------%
\end{document}